\documentclass{article}

\usepackage{arxiv}

\usepackage{epsfig}

\begin{document}

\title{An introduction to distributed training of deep neural networks for segmentation tasks with large seismic datasets}

\renewcommand{\thefootnote}{\fnsymbol{footnote}}

\author{
  Claire Birnie \\
  KAUST, KSA \\
  formerly Equinor ASA, Norway\\
  \texttt{cebirnie@gmail.com} \\
   \And
  Haithem Jarraya \\
  Jarraya Consultancy Ltd.\\
  London, United Kingdom \\
   \And
  Fredrik Hansteen \\
  Equinor ASA\\
  Bergen, Norway \\}

\maketitle

\chead{Distributed NN training with seismic data}

\begin{abstract}

Deep learning applications are drastically progressing in seismic processing and interpretation tasks. However, the majority of approaches subsample data volumes and restrict model sizes to minimise computational requirements. Subsampling the data risks losing vital spatio-temporal information which could aid training whilst restricting model sizes can impact model performance, or in some extreme cases, renders more complicated tasks such as segmentation impossible. This paper illustrates how to tackle the two main issues of training of large neural networks: memory limitations and impracticably large training times. Typically, training data is preloaded into memory prior to training, a particular challenge for seismic applications where data is typically four times larger than that used for standard image processing tasks (float32 \textit{vs.} uint8). Using a microseismic use case, we illustrate how over 750GB of data can be used to train a model by using a data generator approach which only stores in memory the data required for that training batch. Furthermore, efficient training over large models is illustrated through the training of a 7-layer UNet with input data dimensions of 4096$\times$4096 ($\sim7.8$M parameters). Through a batch-splitting distributed training approach, training times are reduced by a factor of four. The combination of data generators and distributed training removes any necessity of data subsampling or restriction of neural network sizes, offering the opportunity of utilisation of larger networks, higher-resolution input data or moving from 2D to 3D problem spaces.

\end{abstract}

\section{Introduction}
The use of Deep Learning (DL) has seen a resurgence in its application to geophysical problems over the past decade. Last century's investigations into the potential benefits of DL methodologies were hampered by technological limitations \cite{dean2018}. Nowadays, access to reasonably powerful compute is freely available with certain cloud providers even offering free GPU provisions in their experimentation environment, for example CoLab. Alongside this, ``tech giants" have open-sourced deep-learning packages en masse, such as Google's TensorFlow package \cite{abadi2016} and Facebook's PyTorch package \cite{paszke2019}. These advancements have significantly lowered the bar for incorporating DL approaches into research projects and, as such, have contributed to the surge in development of deep learning applications for the geoscience domain. Furthermore, training data and pretrained models have become increasingly more available.

Whilst DL methodologies have seen a resurgence across all fields of seismology, and wider geoscience applications, the use of computer vision procedures in particular has been shown to be incredibly useful for seismic processing and interpretation problems, where the `input' data can be treated as an image. \cite{kaur2020} illustrated the use of Cycle Generative Adverserial Networks for groundroll suppression in land seismic data whilst \cite{yu2019} illustrated the potential of Convolutional NNs (CNN) for seismic denoising of random and linear noise signals, as well as multiple suppression. The use of Neural Networks (NNs) for interpretation of seismic cubes has been extensively tested over the last five years with promising results being offered from many different approaches varying in both preprocessing, NN architecture and postprocessing. For example, \cite{hami2017} investigate the use of a growing NN for an unsupervised clustering procedure to accelerate seismic interpretation, whilst \cite{wu2018a} illustrated the use of a CNN for identification of faults within a 2D window from a seismic volume.

DL is not just making waves in the active seismic community, it has also begun making headway in passive seismic applications through the introduction of new, more reliable procedures for event detection. From a single station viewpoint, i.e., where traces are handled independently of one another, Recurrent NNs have been shown to be particularly powerful in offering an alternative to the commonly used short-time average, long-time average detection procedure, for example \cite{zheng2018,birnie2020inreview}. Whilst from an array point of view, both \cite{stork2020} and \cite{consolvo2020} have illustrated how CNNs can be used for detecting an events arrival within a certain time-space bounding box.

Despite great advancements being made on tailoring NN architectures for geophysical applications, one large drawback remains: training of large NNs is memory and time expensive. As such, the majority of deep learning applications for seismic datasets require subsampling of the data \cite{alwon2018}. A solution to this is to train NNs in a distributed manner. Using passive monitoring as a use case, this paper walks through the design, implementation and deployment of a deep learning problem that leverages on the ability to distribute the NN training, allowing efficient training of a large NN ($\sim7.8$M trainable parameters) with a large ($>750$GB) training dataset.

\section{Dataset}
\begin{figure}
  \centering
  \includegraphics[width=0.8\textwidth]{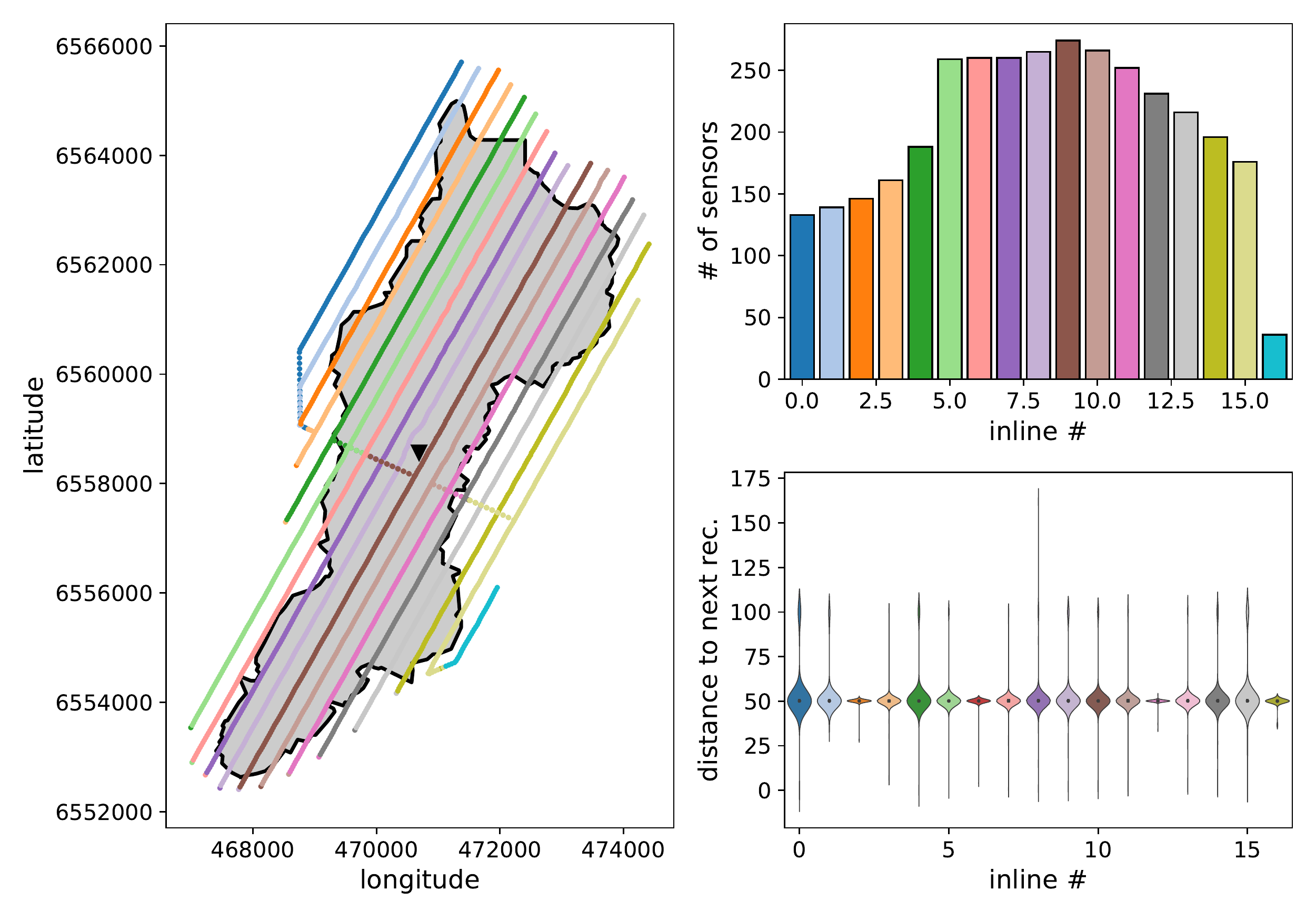}
  \caption{Array information for the permanent reservoir monitoring system deployed over the Grane field in the Norwegian North Sea. The array is separated into "inline" sections as represented by the colourscale. (a) illustrates the array geometry overlaid on the field's polygon with the black triangle indicating the location of the platform. (b) Details the number of sensors per line, whilst (c) details the distribution in distances between neighbouring sensors per "inline".}
  \label{fig:array}
\end{figure}

Similar to the development of many processing, imaging and inversion algorithms, in this study our approach is developed on synthetic data and tested on a field dataset. The field dataset comes from a PRM system deployed on the seabed at the Grane field in the Norwegian sector of the North Sea. The PRM system consists of 3458 sensors, 3-component geophones with a hydrophone, arranged in a psuedo-gridded-style with a sparser ``crossline" backbone as illustrated in Figure \ref{fig:array}. The receiver spacing is approximately 50m along the cables (inline) and 300m between the cables (crossline). Continuously recording at a 500Hz sampling rate, almost 2.4TB of passive seismic data are collected every day.

The system is primarily used for reservoir and overburden monitoring with active seismic surveys. However, it has also been shown to provide invaluable additional information by using it for passive monitoring. For example, drill bit localisation during drilling campaigns \cite{houbiers2020} and interferometric velocity modelling \cite{zhang2019}.

\begin{figure}
  \centering
  \includegraphics[width=0.7\textwidth]{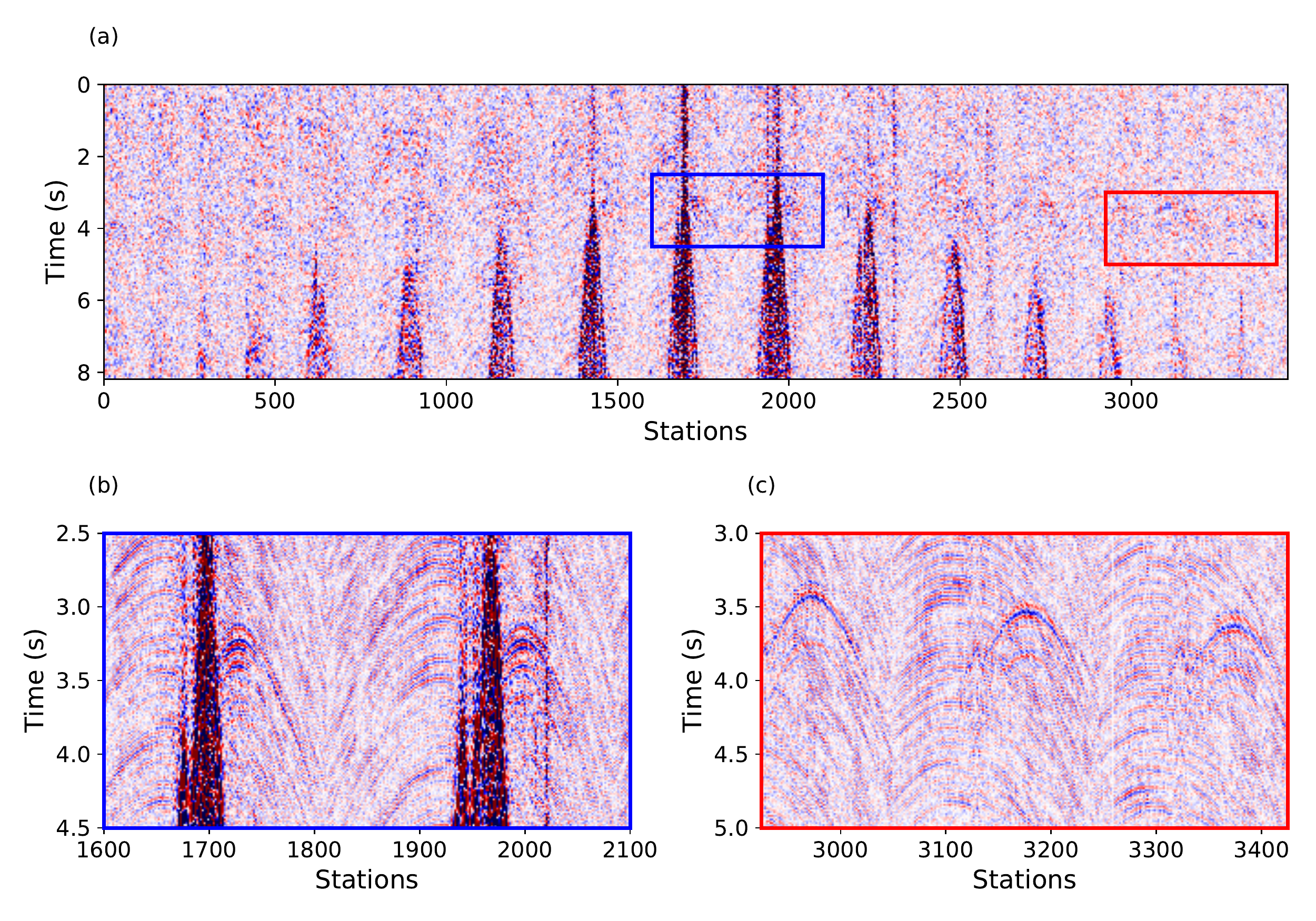}
  \caption{Bandpassed data of the G8 event recorder over the full PRM array (a). The blue box in the center corresponds to the zoomed in data segment shown in (b) highlighting the event arrival at the same time as an onset of platform noise. The red box corresponds to the zoomed in data segment shown in (c) from a quieter section of the array.}
  \label{fig:g8_raw}
\end{figure}

To-date no seismic events have been recorded due to subsurface movement. However, in the summer of 2015 during a drilling campaign, energy waves resulting from a liner collapse were captured in the seismic data. An in-depth analysis of this event was performed by \cite{bussat2018} using a subset of the receivers. The z-component of this event, hereinafter referred to as the G8-event, is illustrated in Figure \ref{fig:g8_raw} and used in this study for the benchmarking of the developed ML detection procedure. 

\section{Methodology}
Defining a clear problem statement is fundamental for the development of any new algorithm, whether ML related or not. For the passive monitoring scenario the problem statement we investigate in this paper is how to develop a \textbf{real-time} event detection procedure that \textbf{utilises the full array}. Two other key elements in the development of ML approaches include: the training dataset and the model architecture. Below we discuss in detail how the problem is set up, how training data is chosen and how the model architecture is adapted for the use case.

\subsection{Solution design}
For the microseismic scenario, events are typically below a SNR of one and therefore a lot of standard processing measures leverage the additional spatio-temporal information that can be captured by using array processing procedures as opposed to trace-by-trace methods. For example, there are a number of different stacking procedures that have been shown to improve detection procedures by increasing the SNR, such as envelope stacking \cite{gharti2010} or semblance stacking \cite{Chambers2010}.

\begin{figure}
  \centering
  \includegraphics[width=1.\textwidth]{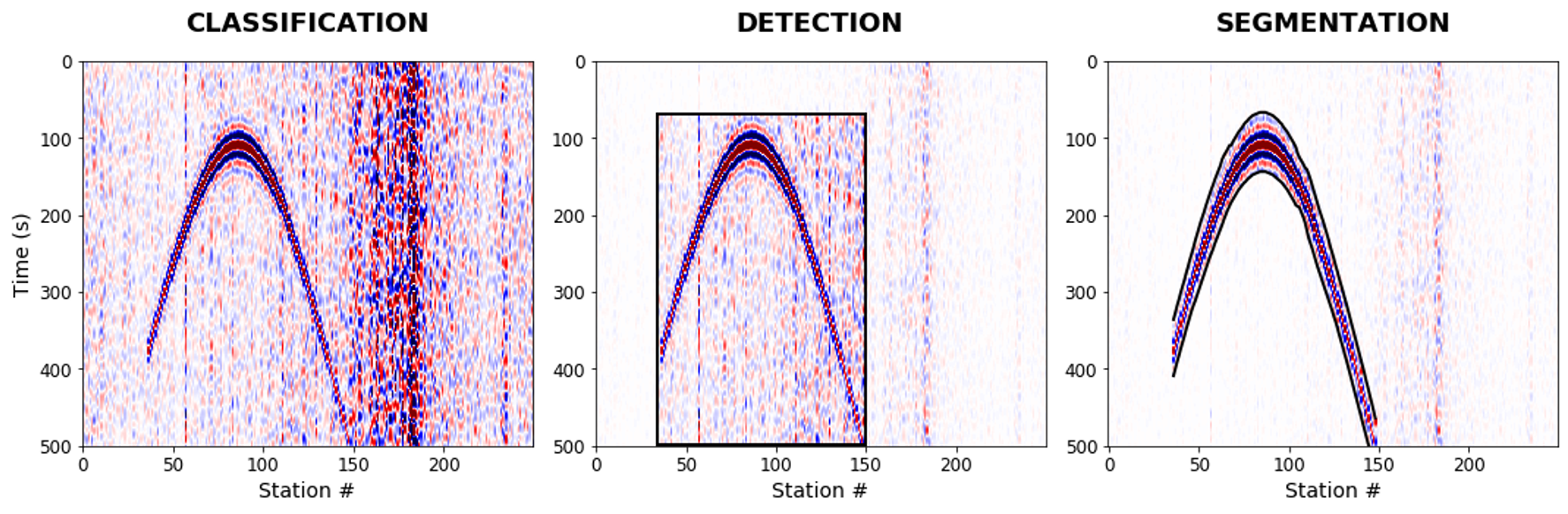}
  \caption{Schematic illustrating how microseismic event detection can be considered as a computer vision task, either as full image classification, object detection, or image segmentation.}
  \label{fig:CV_for_ms}
\end{figure}

Figure \ref{fig:CV_for_ms} illustrates how microseismic event detection can be considered as a computer vision task, whether as a classification, object detection, or image segmentation task. Considering the full array, the identification of the signal within a certain time window can be considered as an image segmentation tasks where each pixel represents a single point in time, $t$, and space, $x$. Therefore the task is to determine for each pixel in the image whether it contains a seismic event or not, i.e., a binary classification per pixel.

\begin{figure}
  \centering
  \includegraphics[width=1.\textwidth]{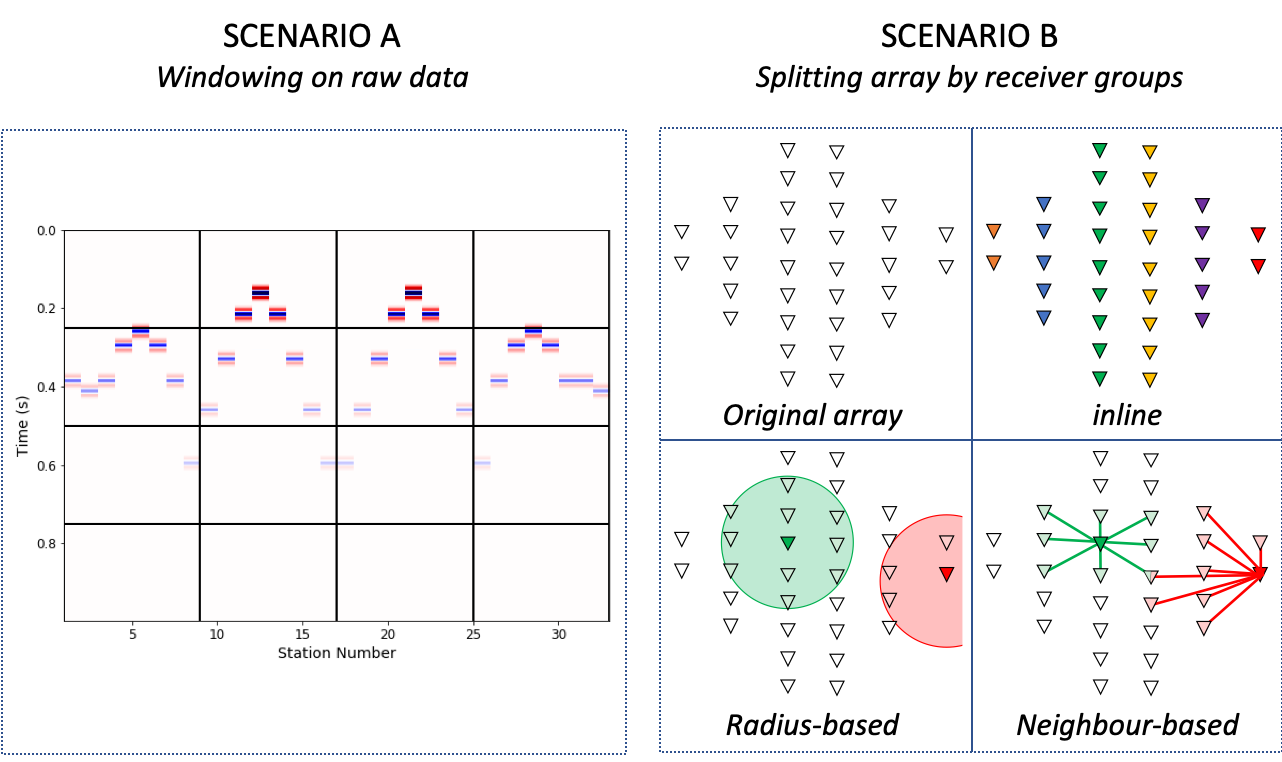}
  \caption{Schematic illustrating possible approaches to windowing of the seismic data prior to developing DL models. }
  \label{fig:windowing_options}
\end{figure}

Sliding window approaches have proved very popular in previous image segmentation tasks on post-stack seismic data. This works particularly well due to the uniform sampling in processed seismic sections from active acquisitions meaning that all windows whether 2D or 3D maintain the same distance between samples. However, this is not the case when working with pre-migrated data, as is often the case in passive monitoring. Figure \ref{fig:windowing_options} Scenario A offers an impression of how a rudimentary, spatio-temporal windowing procedure, the most commonly applied in seismic DL applications, could be implemented for raw passive data on a pseudo-gridded-geometry analogous to the Grane geometry. For this approach, one must only consider/optimise the number of stations to include in the window and the time range on which to span. However, due to the irregular spacing between receivers, there is little consistency in the relationship between event arrivals across the different windows.

Scenario B offers a number of more sophisticated alternatives to Scenario A. As illustrated in Figure \ref{fig:windowing_options}, receivers groups could be selected multiple ways: by inline grouping, a radius-based approach from central receivers or a nearest-neighbour approach. A number of design decisions must be considered with these approaches: the number of receivers per group; the number of models to be created (e.g., one per group); how to handle over-utilisation of receivers where they are grouped into multiple groups; as well as the obvious, which grouping method to use. For the inline and radius-based approaches the number of stations would change between each window therefore requiring different NN models per group. Fixing the number of `neighbours', as illustrated by the neighbour-based approach, would remove the complication of varying input dimensions however would still introduce inconsistencies in the spatial distribution of arrivals, particularly at the edges of the array.

The alternative to splitting the data is to develop an image segmentation procedure that uses all 3458 sensors simultaneously. This removes the complications of determining the optimal receiver groupings (and number of models), however it introduces computational complexities due to the size of each data ``observation". To provide a comparison, most imaging recognition tasks utilise input dimensions of 256$\times$256 \cite{deng2009}. Other DL applications on seismic data have ranged from input windows of 24$\times$24 \cite{ma2018} to 100$\times$100 \cite{guo2018} to 128$\times$128$\times$128 \cite{wu2019}. $3458$ is substantially larger than most input dimensions, as such the remainder of the paper will focus on how to efficiently train NNs with large input dimensions.

\subsection{Data creation}
In the seismic space there are three main options for gathering training data: field data collection, laboratory created data,  or synthetically generated  data. A good training dataset must have a large volume of data available, be similar to the data onto which the trained model will be applied and be simple to label. Largely to avoid the tedious annotation procedure typically associated with supervised learning approaches, in this study synthetic datasets were generated for training the model. Historically synthetic datasets have been heavily utilised in the development and benchmarking procedures of new algorithms, and the importance of using realistic synthetics to accurately depict how an algorithm will perform on field data cannot be overstated \cite{birnie2020}. Similarly, to train an ML model that is robust for application to field data, the training data must provide an efficient representation of the variety of waveforms and noises that exist in such recordings however at a reasonable creation speed. In this section we discuss how we have generated a diverse dataset of realistic synthetic seismic recordings for training and evaluation purposes.

\begin{figure}
  \centering
  \includegraphics[width=1.\textwidth]{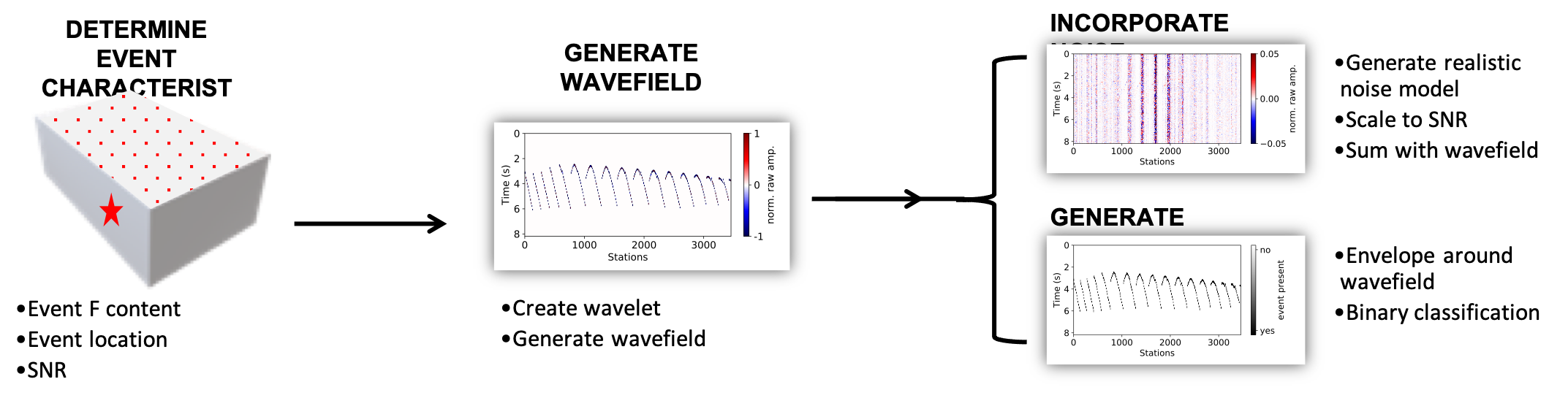}
  \caption{Workflow of the generation and labelling of synthetic data.}
  \label{fig:synthgeneration}
\end{figure}

Using travel times and the standard convolutional modelling approach, synthetic datasets are generated using the workflow as illustrated in Figure \ref{fig:synthgeneration}. First, the source location is randomly selected from a cube in the subsurface centered around the top of the reservoir. The source parameters: wavelet type, frequency content, and SNR are also randomly selected. The wavelet is then generated and the wavefield data is created via convolutional modelling with a scaler accounting for amplitude decay due to geometrical spreading.

Noise is an ever-persistent challenge in seismic field data handling. To make the synthetics representative of field data, synthetic coloured noise models are generated using statistics observed from previously collected passive recordings. The frequency spectrum of the recordings are grouped into 5Hz bands representing the percent of total energy within each band. This is used to scale the coloured noise model such that it has a similar frequency content to recorded noise, similar to the approach of \cite{Pearce1977}. The coloured noise model is then scaled spatially to represent the spatial distribution in energy as typically observed on the array, e.g., higher amplitudes around the vicinity of the production platform.

As well as forming the base of the synthetic seismic dataset, the wavefield data is used to generate the matching ``label" dataset for training and evaluation purposes. As event detection is a binary classification, the labels are either zero or one where one indicates that a wavefield of interest is present. An event's arrival is classified anywhere where the wavefield energy is greater than a specified amount depending on the wavelet type and frequency content. 

For simplifying experimentation of the NN architecture, to be discussed below, the length of each synthetic dataset is 4096 time samples which equates to 8.192 seconds, given a 2ms sampling frequency. Assuming the energy bands for noise spectrum and the array geometry are preloaded, it takes 1.7 seconds from start to end of the generation procedure of a single data sample (when computed on a 2.9GHz, 6-core Intel Core i9 machine with 32GB RAM).

\subsection{Model architecture}
\begin{figure}
  \centering
  \includegraphics[width=1.\textwidth]{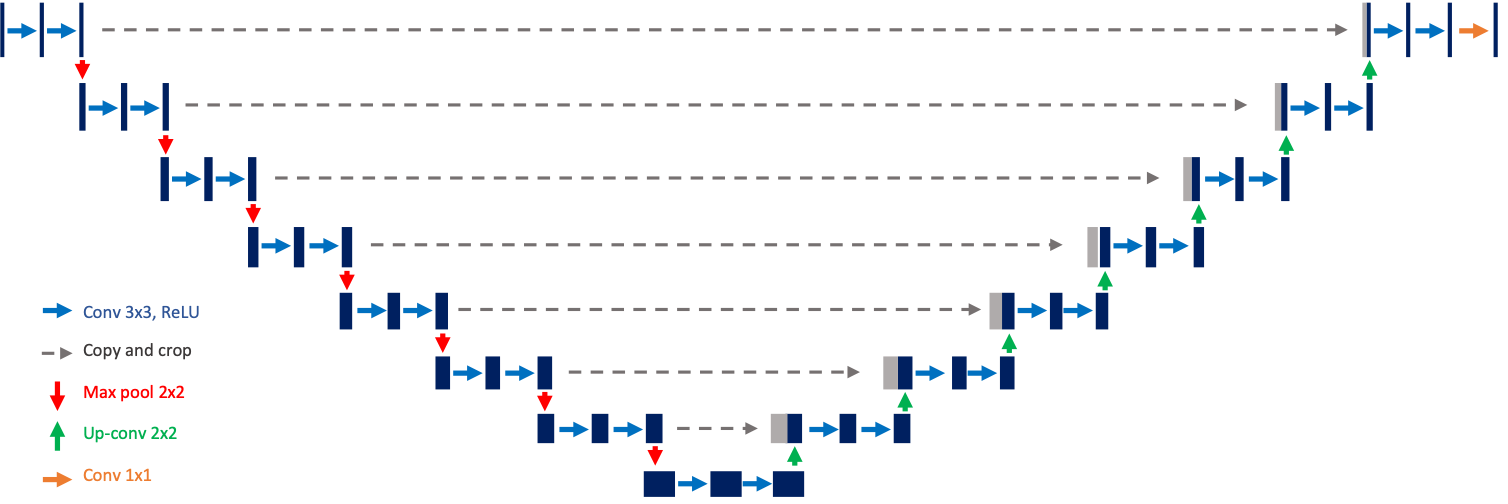}
  \caption{Seven layer UNet architecture.}
  \label{fig:model_architecture}
\end{figure}

The U-Net architecture of \cite{ronneberger2015} has become the workhorse for most image segmentation tasks on seismic data, following on from its successful application for image segmentation in medical imaging. The standard U-Net architecture follows the form of a contracting (left) path and an expansive (right) path as illustrated in Figure \ref{fig:model_architecture}. The contracting path has the ability to capture context and consists of repeated blocks of: two 3$\times$3 convolutions each followed by a rectified linear unit (ReLU) and a 2$\times$2 max pooling operation with stride 2 for downsampling. Whilst the expansive path enables precise localization and consists of repeat blocks of: two 3$\times$3 convolutions each followed by a rectified linear unit (ReLU) and a 2$\times$2 upsampling convolutional layer with a stride of 2.

As noted by \cite{ronneberger2015} in the original U-Net study: \textit{ ``To allow a seamless tiling of the output segmentation map ..., it is important to select the input tile size such that all 2x2 max-pooling operations are applied to a layer with an even x- and y-size."}. 3458, the number of sensors in the Grane PRM system, when halfed becomes an odd number, $1729$, therefore it is not possible to make a U-Net without altering the input dimensions. An additional 638 null traces were added to the array such that the input dimension became 4096 - a binary number meaning  that we can divide by two all the way down to one. These input images are now orders of magnitude larger than \cite{ronneberger2015}'s study, whose experiment used images of 512$\times$512 pixels. 

In the original U-Net study, four layers were utilised, reducing the data dimensionality down to 32 at the base of the NN. For the Grane example, an additional three layers are required to reduce the data down to the same dimensions. For the convolution steps we begin with four filters at the top layer, multiplying by a factor of two at each reduction step. The incorporation of the additional layers and following the filter methodology, the resulting model has $\sim7.8$M number of trainable parameters.

\section{Implementation}
The large dimensions of the data are not the only ``size" complexity arising in this use case due to the data types which are involved. Typically images are stored with a data type of \textit{uint8} whilst seismic data is stored with a \textit{float32} data type. Therefore, a seismic section with the same dimensions as an image is four times larger, impacting memory requirements for NN training. This complexity presents a challenge when loading data into memory for  training the NN. For the majority of image segmentation tasks the full training set is loaded into memory prior to training. In this experiment, each labelled seismic section is 108 MB therefore it is not feasible to load $6000+$ into memory. 

TensorFlow's dataset functions offer a manageable solution to the memory limitation challenges encountered due to the datasize. This allowed the storing of only the required data samples per step, therefore removing any necessity to reduce the size of the model or the input data dimensions.

In the data creation section above we argued for the use of synthetic datasets for training purposes. However, there are two approaches to how this can be implemented. Firstly data can be pre-made, written to file and read in as needed. Alternatively a data generator can be implemented that creates data on-the-fly. For this specific use case, we calculated that it would take $\sim4$ hours and $756$GB of storage for the first option, additionally taking 2 seconds per file to be read in -  assuming the data is stored as a TensorFlow Tensor. The second option has the advantage that no additional storage is required however the data would need to be re-generated every epoch. In this case, the generation time is similar to the loading time and as such there is little difference in the processing time of either approach (considering only the reading time for the first approach). Therefore, due to the lowered storage requirements, we choose to implement the second approach of generating the data on-the-fly. The data generator was seeded with the sample number such that the same data was generated per epoch and could be replicated at any future point.

\section{Training}
The model has $\sim7.8$M trainable parameters with 6000 seismic sections per epoch with an additional 1000 samples generated for validation. Using a single machine with a large GPU \footnote{6 core, 112GiB, 1xNvidia V100 GPU}, a single training sample takes $\sim28$s. Therefore for one epoch, excluding validation, on a single GPU machine takes $\sim47$hours. 

\begin{figure}
  \centering
  \includegraphics[width=1.\textwidth]{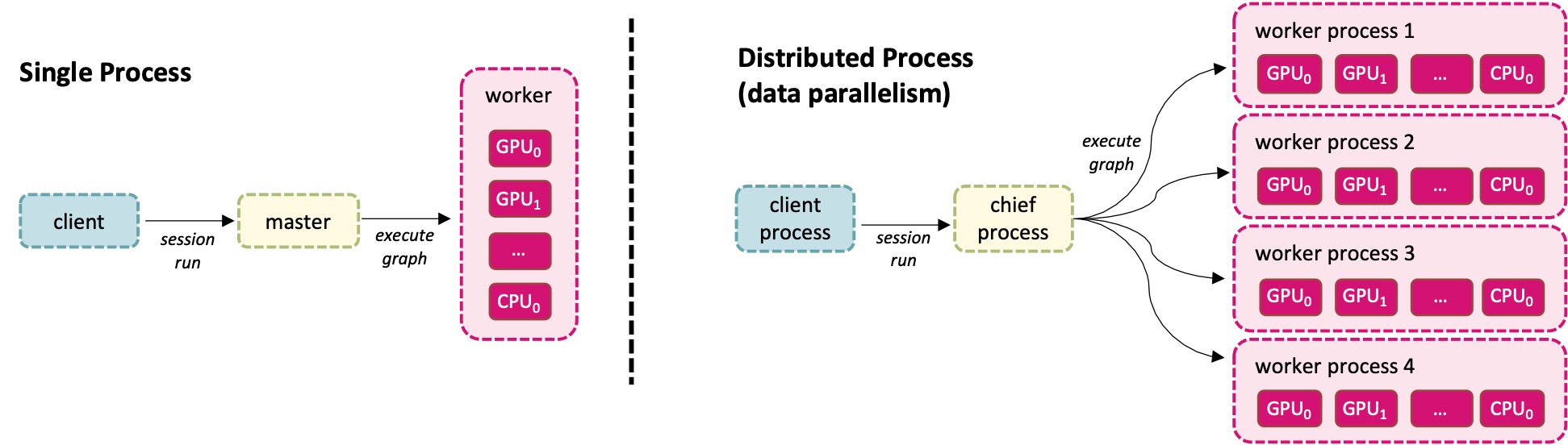}
  \caption{Comparison between a single process for training vs a distributed process using a data parallelism strategy with 4 workers. The evaluator node is not illustrated.}
  \label{fig:dist_strategy}
\end{figure}

Parallelisation of the training regime can drastically decrease the total training time and is a functionality available in both the two biggest machine learning Python libraries: TensorFlow and PyTorch. In this example, we use a batch-splitting (data parallelism) approach implemented by using TensorFlow Estimators with 4 workers as illustrated in Figure \ref{fig:dist_strategy}. A separate evaluator node is also added to our resource pool such that training is not paused during the validation steps. We follow a synchronous updating procedure requiring each worker to complete its batch and return weight updates to the chief before workers can begin on the next batch of training samples. Utilising 4 workers of the same specs as the GPU machine in the serial example, with an additional evaluator node, training time for one epoch is reduced to under $12$ hours. Note, some additional compute time is introduced due to both communication and waiting (due to the synchronous training mode).

The training is run on cloud resources and orchestrated using Kubernetes. The training scripts were written and tested locally on small, dummy datasets before being incorporated into a custom Docker Image. A cluster of cloud compute resources were commissioned, in this case five GPU machines with the specs as previously described. A fileshare was mounted to the resources containing the necessary files for the synthetic data creation - geometry and noise energy frequency bands - allowing access to the files as if they were locally stored. Distributed training is initialised via applying a Kubernetes \textit{.yaml} file to the cluster. The \textit{.yaml} contains all the necessary information regarding file paths, number of resources to use for training and validation, as well as the additional Python inputs such as number of training samples per epoch, snapshot frequency, range of synthetic parameters, etc. Once the Kubernetes job has been initiated, the required number of pods are created, in our case one chief, three additional training pods and an evaluator, and the training job begins.

\begin{figure}
  \centering
  \includegraphics[width=1.\textwidth]{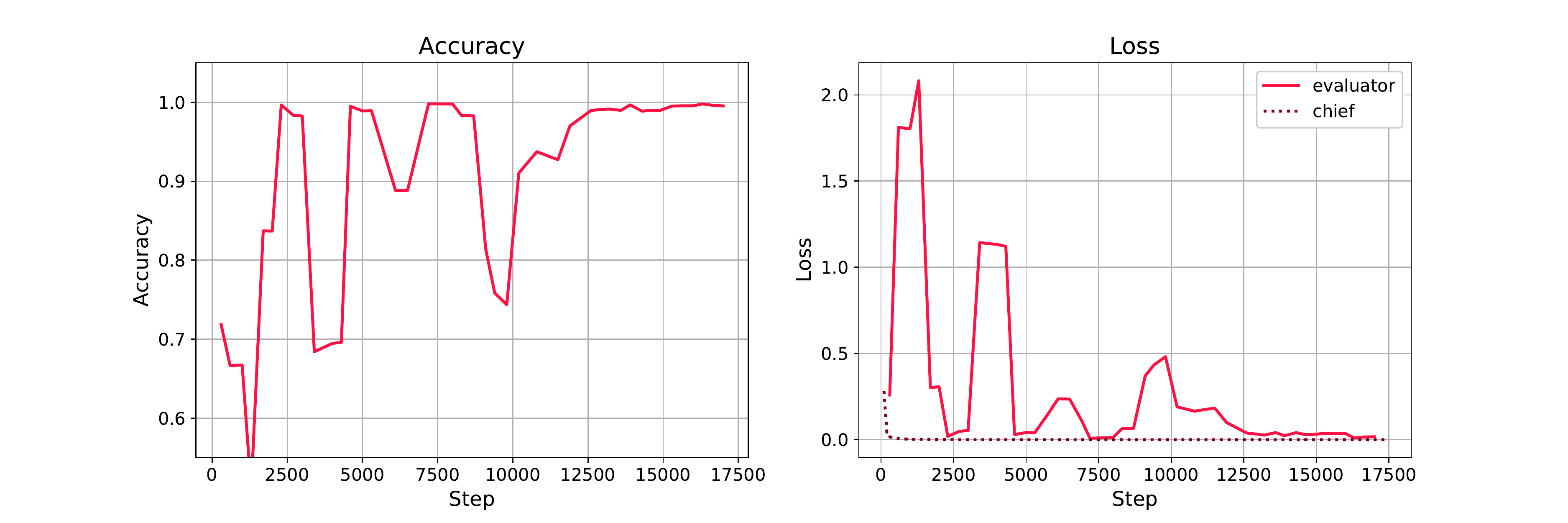}
  \caption{Progression of the model accuracy and loss during training.}
  \label{fig:losses}
\end{figure}

The model is saved at every checkpoint during the training procedure allowing analysis of the model whilst training is ongoing. Training ran for approximately 6 days, covering 12 epochs (i.e., 18000 training steps of 4 samples each), before the model was deemed sufficiently trained via a qualitative analysis of detection performed on newly created (i.e., blind) synthetic recordings. Figure \ref{fig:losses} illustrates the progression of the model's accuracy and loss with respect to the evaluation dataset, as well as the chief's loss, over the training period.

\subsection{Evaluation}
\begin{figure}
  \centering
  \includegraphics[width=0.9\textwidth]{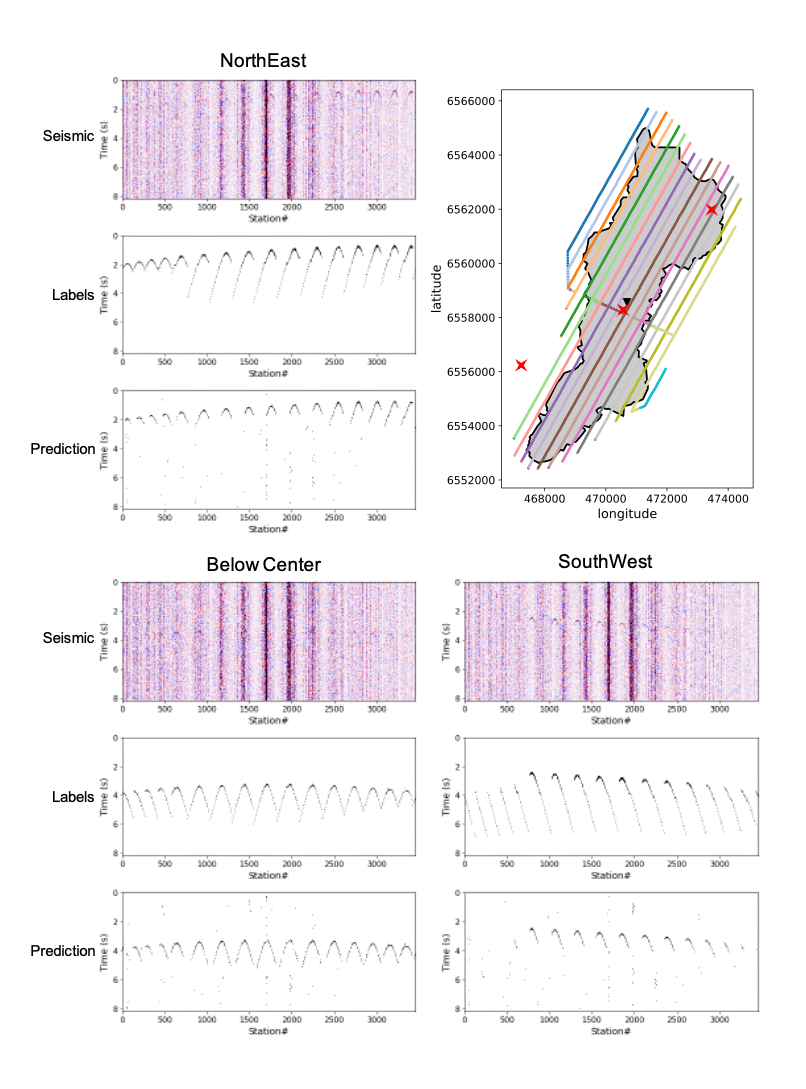}
  \caption{UNet detection's on synthetic seismic events with source origins in different subsurface locations: NorthEast of the array, below the center of the array and to the SouthWest of the array, as illustrated by red crosses in the array map. The top panel shows the synthetic data, the middle panel shows the labels corresponding to the synthetic and the bottom panel shows the UNet's detection.}
  \label{fig:synth_varying_loc}
\end{figure}

\begin{figure}
  \centering
  \includegraphics[width=0.8\textwidth]{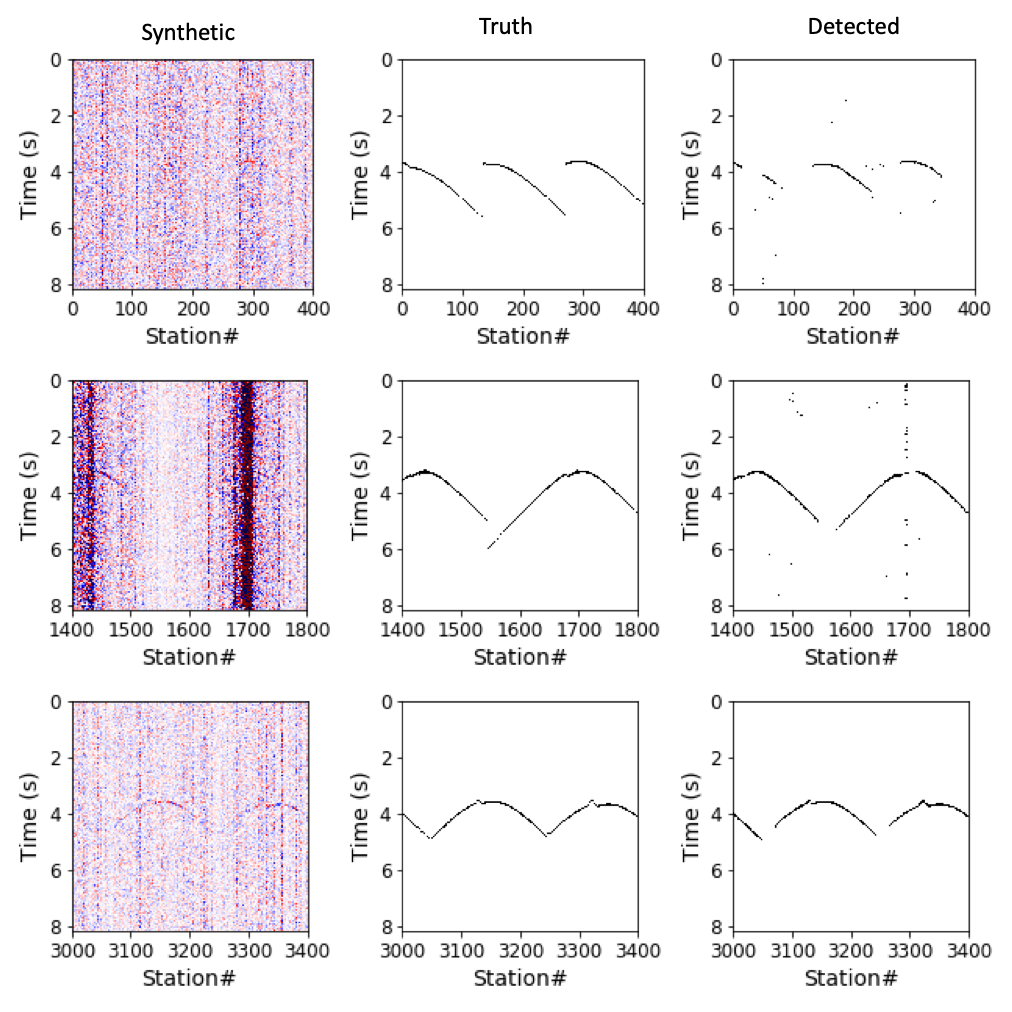}
  \caption{Magnified results from the event below the center of the array as illustrated in \ref{fig:synth_varying_loc}. The first row comes from receiver lines in the West of the array, the middle row includes the two inlines closest to the platform in the center of the array, and the bottom row includes receivers in the furthest East lines.}
  \label{fig:synth_varying_loc_zoomed}
\end{figure}

Once sufficiently trained, a number of new synthetic datasets were generated that the model was not exposed to during the training period covering a range of different event locations. Figure \ref{fig:synth_varying_loc} illustrates the performance of the trained network on predicting the event arrival for three events of the same magnitude (SNR=0.4), one to the NorthEast of the array, one below the center of the array, and one to the SouthWest. Whilst the moveout patterns are significantly different the network manages to accurately detect the arrivals. Figure \ref{fig:synth_varying_loc_zoomed} zooms in on the recordings from different sections of the receiver array, illustrating how the detection procedure accurately handles the varying amplitude of arrivals across the array as well as the varying local moveouts. In both Figures \ref{fig:synth_varying_loc} and \ref{fig:synth_varying_loc_zoomed}, there is little-to-no additional noise in the detection arising from the heightened noise levels around the platform site.

\begin{figure}
  \centering
  \includegraphics[width=0.9\textwidth]{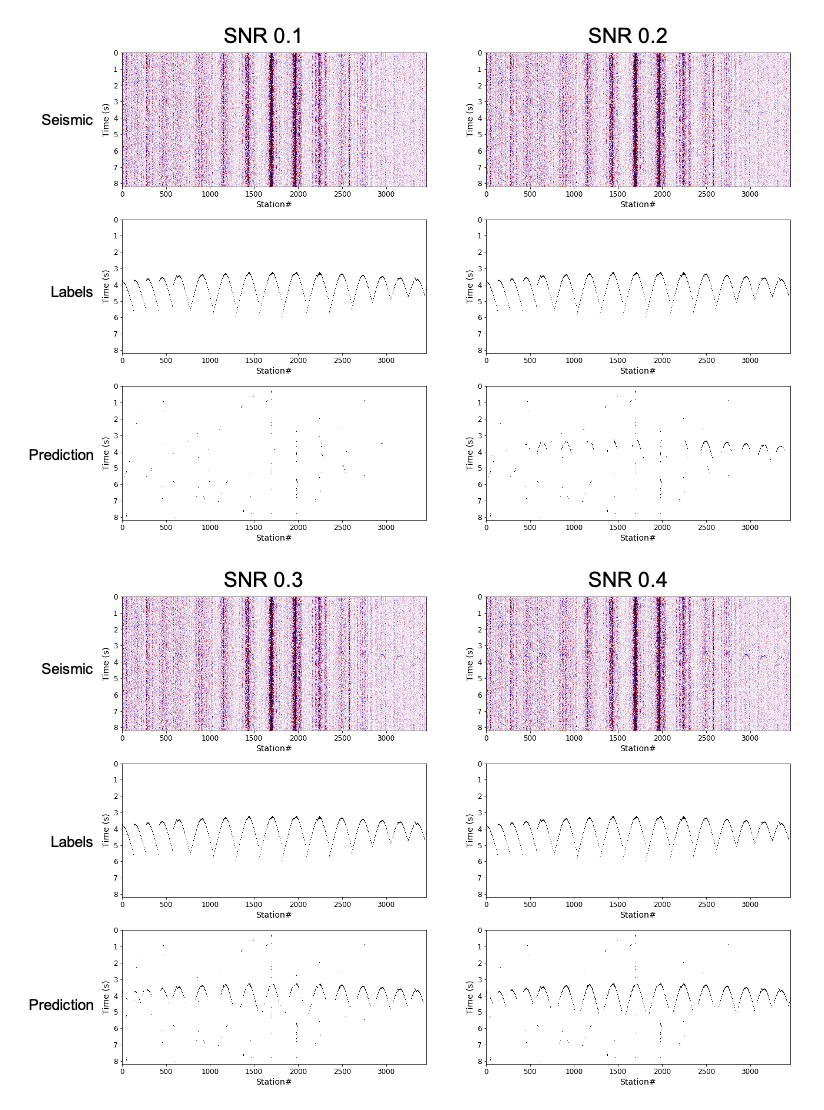}
  \caption{SNR investigation on the performance of the trained UNet with the event always originating from the same subsurface location.}
  \label{fig:synth_varying_snr}
\end{figure}

A similar analysis is run analysing the sensitivity of the trained segmentation model to varying SNRs. Figure \ref{fig:synth_varying_snr} illustrates how the detection procedure can handle low SNR events. As expected, decreasing the SNR of arrivals results in increasing noise in the detection procedure. Down to an SNR of 0.2 the arrival shape is clearly visible within the prediction section however at an SNR of 0.1 the event arrivals are no longer easily identifiable.

\begin{figure}
  \centering
  \includegraphics[width=1.\textwidth]{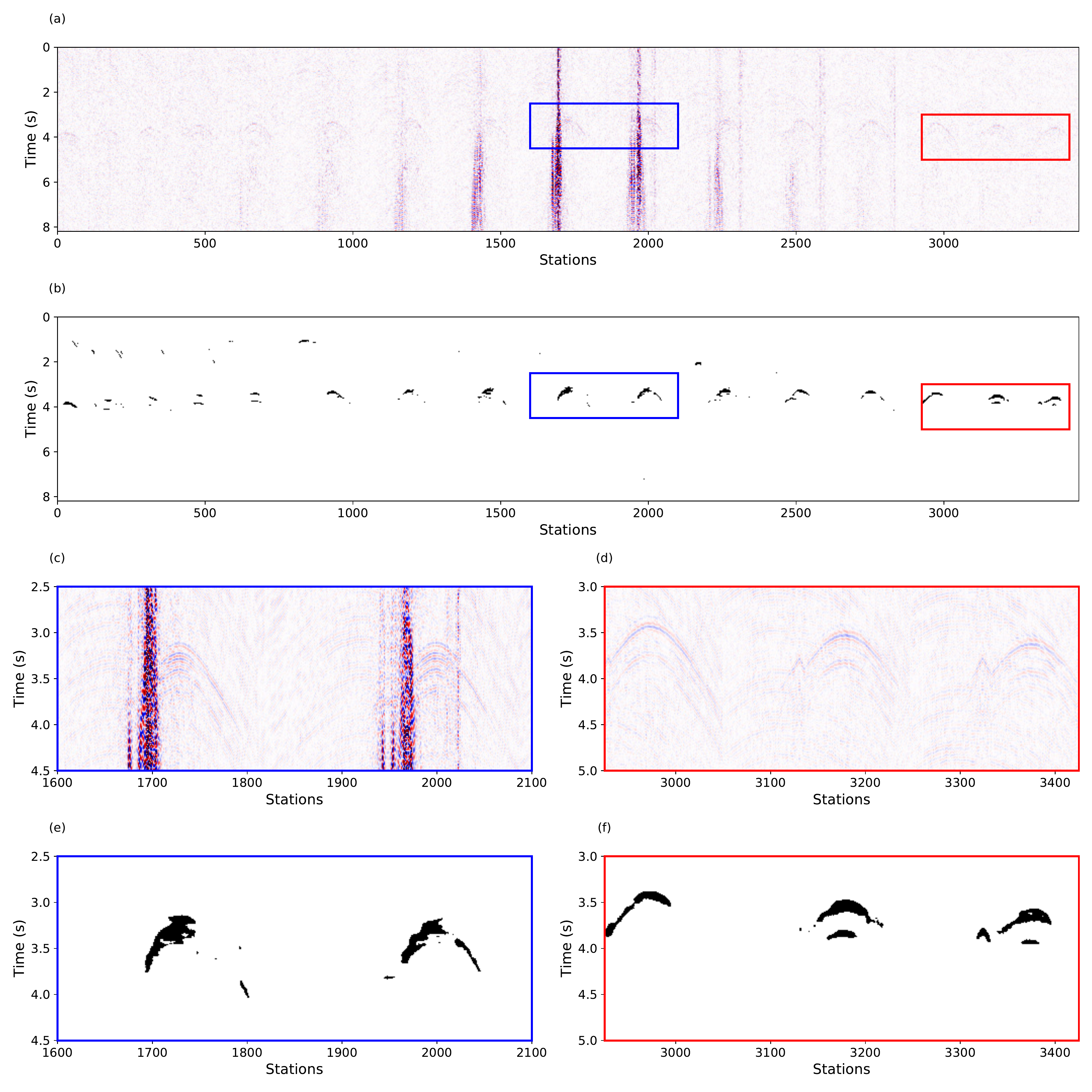}
  \caption{UNet event detection on the Grane G8 liner collapse event. The blue box in the center corresponds to the zoomed in data segment and detections shown in the bottom left column highlighting the event arrival at the same time as an onset of platform noise. The red box corresponds to the zoomed in data segment and detections shown in the bottom right column from a quieter section of the array.}
  \label{fig:g8}
\end{figure}

To ensure the trained model is applicable to field data, it is applied to the previously described G8 event. Figure \ref{fig:g8} displays the 8 second seismic recording with the event alongside the UNet predictions. The blue box highlights the arrival on a particularly noisy receiver grouping whilst the red box indicates the arrival on a quieter group of receivers at the edge of the array. The event is clearly detected across the majority of receivers without detecting the platform noise that begins halfway through the recording.

\section{Discussion}
The aim of this study was to investigate possible methodologies for the training and application of large deep NNs on seismic datasets without the requirement of subsampling or windowing. The ability to handle larger input data dimensions as well as train larger models offers the opportunity for capturing additional spatio-temporal information from the seismic data - a well documented approach for enhancing SNR. The solution design section of this paper, in particular Figure \ref{fig:windowing_options}, highlighted the complications in developing a generic model to be applied on either receiver lines or by windowing the array -  for this specific use case. As such, the simplest path was to process the full array in one go and leverage technological advances to allow the training of such a large model.  

There also exist a number of other use cases which naturally permit windowing however that may benefit from using larger windows. Fault detection is one such task that is often reduced to a 2D problem despite the `original' 3D subsurface data volume. For example, \cite{guo2018} extract 2D slices from a 3D seismic cube, explicitly stating: ``The dimension reduction from 3D to 2D is to reduce the time to train the CNN." Similarly, \cite{ma2018} provide 24$\times$24 images with an inline, crossline and time section as input channels to a 2D NN, rendering the problem psuedo-3D. However, they stop short of utilising a full 3D input. Distributing the training allows the possibility of using larger input data dimensions (increasing window sizes or adding an additional dimension) therefore, either capturing a larger spatio-temporal area or offering the opportunity to use higher resolution data.

A trade-off can occur between input data dimensions and model size where, as opposed to subsampling data, a smaller, simpler model is used. For example, in the microseismic event detection use case both \cite{stork2020} and \cite{consolvo2020} have trained CNNs to detect a time-space box in which an arrival is detected. The smaller computational requirement allows for a faster training procedure however less information can be derived from the models' predictions. For object detection the returned information is that of a bounding box with the same ``arrival time" for all receivers as opposed to segmentation procedures which detect arrival times per trace. \cite{wu2019} provide another example of where a smaller network has been utilised. They used a simplified UNet with a reduced number of layers for a 3D fault detection procedure which allowed for significant ``savings in GPU memory and computational time". The procedure for efficient training detailed in this paper provides the opportunity to increase model dimensions whilst still keeping a reasonable training time.

It should be noted here that not all deep learning applications on seismic data require subsampling. For example, should the same segmentation procedure developed in this paper be adapted for a different, smaller permanent array, such as the 50 receiver array at Aquistore \cite{stork2018}, the input dimensions would be smaller than those used in the original UNet implementation rendering any discussion on subsampling unnecessary. However, these use cases are becoming rarer, particularly with the adoption of densely sampled fiber optic cables for permanent monitoring.

The success of a model is highly dependent on its training data, and the use of synthetic datasets for training has become common-place in seismic deep learning procedures, for example \cite{huang2017,pham2019,wu2019,cunha2020}. In traditional synthetic data usage for developing and benchmarking algorithms it has been shown that the more realistic the synthetic data the better for understanding uncertainties and identifying pitfalls \cite{birnie2020}. However, there is a trade-off between similarity to field data and computational cost which is particularly applicable when developing the large volume of datasets required for training deep learning models. In this use case, we found that generating the waveform data via a wave propagation procedure would be too computationally expensive for what we classified as a reasonable generation time - sub two seconds. As such, used a simple convolutional modelling procedure incorporating geometric spreading and assuming a homogeneous velocity model for the traveltime computations. Similarly, for the incorporation of noise in the dataset, the generation of realistic noise models (i.e., non-stationary, non-Gaussian, non-white noise), such as via a covariance-based approach \cite{Birnie2016}, was deemed too timely. Therefore, an approach similar to that of \cite{Pearce1977} is used which generates a stationary noise model that accurately replicates the frequency content of recorded noise. As of yet, an analysis has not been published to show the trade-off between the complexity/reality of synthetics and the performance of the trained model. In this use case the resulting network produces acceptable predictions on this field dataset, but more testing is required to fully assess the model’s performance on a wider variety of field data with varying noise and event properties.

Despite many advancements in detection algorithms over the years, \cite{skoumal2016} highlighted how computational cost is a big barrier preventing the majority of these algorithms making it into a production toolbox. One of the key criteria of such a detection algorithm is its real-time  applicability. Whilst the training took 6 days utilising four GPU machines, detection can be performed in under 3 seconds on a 2.9GHz, 6-core Intel Core i9 machine with 32GB RAM for an 8 second recording segment. Therefore, once trained the model can be used for real-time monitoring applications without any requirement of large computational resources or parallelisation across multiple machines.

\section{Conclusion}
The majority of deep learning applications for seismic data involve the subsampling or windowing of the dataset. In this paper, we illustrate how through the distribution of training, larger networks can be efficiently trained, removing the need for subsampling and/or windowing. Illustrated on a microseismic monitoring use case, the paper walks through the stages of the deep learning project, from synthetic training data creation to adapting a standard model architecture to distributed model training and finally to model evaluation using both synthetic and field datasets. Whilst illustrated on a scenario where data windowing is non-trivial, the benefits of not windowing data, or using larger windows than previously possible, has great potential for other segmentation tasks such as fault and horizon detection.

\section{ACKNOWLEDGMENTS}
The authors would like to thank the Grane license partners Equinor Energy AS, Petoro AS, Vår Energi AS, and ConocoPhillips Skandinavia AS for allowing to present this work. The views and opinions expressed in this abstract are those of the Operator and are not necessarily shared by the license partners. The authors would also like to thank Ahmed Khamassi and Florian Schuchert for their invaluable support on the data science elements of this project, as well as Marianne Houbiers for her insightful discussions on the application of DL for passive monitoring.

\newpage

\bibliographystyle{unsrt}  
\bibliography{bibliography.bib}

\end{document}